\documentstyle[prl,aps]{revtex}
\input epsf.tex

\newcommand{\nc}{\newcommand}
\nc{\beq}{\begin{equation}}
\nc{\eeq}{\end{equation}}
\nc{\beqa}{\begin{eqnarray}}
\nc{\eeqa}{\end{eqnarray}}
\nc{\lra}{\leftrightarrow}
\newcommand{\k}{\kappa}
\def\sfrac#1#2{{\textstyle{#1\over #2}}}
\nc{\sss}{\scriptscriptstyle}
{\nc{\lsim}{\mbox{\raisebox{-.6ex}{~$\stackrel{<}{\sim}$~}}}
{\nc{\gsim}{\mbox{\raisebox{-.6ex}{~$\stackrel{>}{\sim}$~}}}

\def\sqe14{\sqrt{\phantom{iAAA^1}} \!\!\!\!\!\!\!\!\!\!\!\!\!\!\!\!\!\!
   1+\sfrac{4}{\hmo} }

\begin{document}
\twocolumn[\hsize\textwidth\columnwidth\hsize\csname@twocolumnfalse%
\endcsname

\title{No-Go Theorem 
for Horizon-Shielded Self-Tuning
Singularities}

\author{James M.~Cline and Hassan Firouzjahi}

\address{Physics Department, McGill University,\\
3600 University Street, Montr\'eal, Qu\'ebec, Canada H3A 2T8}

\maketitle

\begin{abstract} 

We derive a simple no-go theorem relating to self-tuning solutions to
the cosmological constant for observers on a brane, which rely on a
singularity in an extra dimension.  The theorem shows that it is
impossible to shield the singularity from the brane by a horizon, unless
the positive energy condition ($\rho+p \ge 0$) is violated in the bulk
or on the brane.  The result holds regardless of the kinds of fields
which are introduced in the bulk or on the brane, whether $Z_2$ symmetry
is imposed at the brane, or whether higher derivative terms of the
Gauss-Bonnet form are added to the gravitational part of the action.
However, the no-go theorem can be evaded if the three-brane has spatial
curvature.  We discuss explicit realizations of such solutions which
have both self-tuning and a horizon shielding the singularity.

\vskip0.05in
{PACS: 98.80.Cq \hfill McGill-00-18}
\vskip0.2in
\end{abstract}

]
\section{Introduction}

It is an interesting hypothesis that we might live on a 3-brane embedded
in 5-dimensional Anti-deSitter space \cite{RS}; not only could this idea
solve the hierarchy problem by explaining why mass scales on our brane
are exponentially suppressed compared to $M_{\rm Planck}$, but it might
provide some phenomenological link between string theory, through the
AdS/CFT correspondence, and TeV-scale physics \cite{AHPR}.

At the same time, the braneworld scenario has created the hope of somehow
circumventing Weinberg's no-go theorem for solving the cosmological
constant problem using an adjustment mechanism, by virtue of introducing
an extra dimension.  Some attempts along these lines were made by
\cite{ADKS,KSS}, in which a scalar field in the bulk adjusted itself to
yield a static solution to Einstein's equations, for a range of values of
the brane tension.  These solutions relied upon singular behavior of the
scalar somewhere in the bulk, which was shown by ref.\ \cite{Nilles} to be
simply a way of hiding the fine-tuning problem, since a proper treatment
required insertion of a new brane at the singularity, whose tension must
be tuned with respect to that of the visible brane.\footnote{Moreover the
original self-tuning solutions were shown to require a fine-tuning of the
initial conditions in order to avoid motion of the singularity with
respect to the brane \cite{BCG}.} In essence, the original self-tuning
idea was pretending to gain extra free parameters by no longer requiring
boundary conditions to be satisfied at the boundary of the bulk where the
singularity was appearing, and where a brane would normally have appeared.

A significant attempt to improve on this situation was made in ref.\
\cite{CEG,CEG2}.  Their idea was to render the singularity more physical
by introducing a horizon between it and the visible brane, in the same way
that the Schwarzschild black hole singularity is hidden.  In fact the bulk
geometry is the AdS-Schwarzschild (or AdS-Reissner-Nordstr\"om in the case
of a charged black hole) generalization of AdS, in which a singularity
appears for all of 3-space at some position in the bulk. The singularity
could thus be described as a black brane, though we will follow common
usage and call it a black hole (BH).

The significance of the AdS-Schwarzschild solutions has become apparent
in a number of works that deal with braneworld cosmology.  In ref.
\cite{Kraus} it was shown that this is the bulk solution which gives
rise to the dark radiation term that was shown to be a possible addition
to the Friedmann equation for the expansion of the brane.  Ref.\
\cite{Gubser} subsequently identified the dark radiation as being
identical to the thermal excitations of the CFT degrees of freedom in
the context of the AdS/CFT correspondence.  In ref.\ \cite{HMR} it was
shown that the bulk black hole must form in the early universe, since
gravitational radiation emanating from the hot visible brane becomes
infinitely dense as it falls toward the AdS horizon, for any
cosmologically relevant initial brane temperature.

In contrast to these cosmological solutions, where the brane is moving
away from the BH and thus seeing a bulk which becomes increasingly
AdS-like (or alternatively, the dark radiation term in the brane
Friedmann equation is redshifting away), ref.\ \cite{CEG} finds a class
of static solutions, so that the effect of the BH on the brane can be
felt at arbitrarily late times.  A very interesting application is that
Lorentz invariance is broken, and gravitational signals travel with
an average speed different from that of light on the brane.  Most
germane for the present work is that ref.\ \cite{CEG} also finds
self-tuning solutions with a horizon in the case of the charged (RN)
black hole, where the mass and the charge of the BH adjusts itself to
the energy density $\rho$ on the brane; but self-tuning and the horizon
can coexist only if the positive energy condition is violated on the
brane: $\rho < -p$.

Our motivation for the present work was to try to remove this seemingly
unphysical restriction on the solutions and allow for positivity of the
brane stress energy tensor.  After various failed attempts we realized
that the Einstein equations can be manipulated to show in a simple way why
it is impossible to improve the situation by adding extra matter fields to
the Lagrangian.  This is our no-go theorem, which is given in the next
section.  In section III we show that the theorem also holds when the
gravitational part of the action is supplemented with a particular
higher-derivative correction, the Gauss-Bonnet term.  In section IV we
discuss a way of evading the theorem: giving positive curvature to the 3-D
spatial hypersurfaces parallel to the brane.  We generalize a solution of
this type which was derived for the AdS-Schwarzschild case to the case of
nonvanishing charge and show the range of parameters where self-tuning and
a horizon coexist.  In section V we show that it is not possible to put
the curvature into extra dimensions instead of the usual 3-D space, which
would have been desirable for describing our flat universe.  Conclusions
are given in section VI.

\section{The No-Go Theorem}
We begin with the following general action:
\beq
S = \int d^{\,5}\!x\sqrt{-g}\left(\frac{R}{2\k^{2}}+{\cal L}_{B}\right)
+\int d^{\,4}\!x \sqrt{-g}{\cal L}_{b}
\eeq
where ${\cal L}_{B}$ and ${\cal L}_{b}$ are the Lagrangian densities in
the bulk
and on the brane respectively, and $\kappa^2 = 1/M_5^3$ in terms of the 
5-D Planck mass $M_5$.  The ansatz for the metric, which includes the
AdS, AdS-Schwarzschild or AdS-RN geometries, is

\beq
\label{metric}
ds^2=-h(r)dt^{2}+a(r) d\Sigma_k^{2}+{h(r)}^{-1}dr^{2}
\eeq
where $d\Sigma_k^{2}$ is the line element for a homogeneous 3-D space
of constant spatial curvature with $k=0,\pm 1$.  The 5-D generalization of
Birkhoff's theorem guarantees that this form for the bulk metric is 
a general solution (in the appropriate coordinate system) when there is
only a cosmological constant
\cite{Kraus,Bowcock} or a U(1) gauge field \cite{CEG} in the
bulk.  However we will also take it to be our ansatz for the metric when
there are more general sources of stress-energy in the bulk.  Since we are
interested in static solutions, a coordinate system can always be found
which puts the metric into the form (\ref{metric}).
 For definiteness we will write the 3-D part of the metric as
\beqa
	d\Sigma_k^{2} &=& {dx^2+dy^2+dz^2\over \left(1 + \sfrac14
	k(x^2+y^2+z^2)\right)^2}\nonumber\\
	&\equiv& \Sigma^2_k(x,y,z) ( dx^2+dy^2+dz^2 ).
\eeqa
The nonzero components of the Einstein tensor are
\beqa
G_{00}&=&-\frac{3}{4}h\left(\frac{a'}{a}h'+2h\frac{a^{\prime
\prime}}{a} -4{k\over ah}\right)\nonumber\\
G_{ii}&=& a \Sigma^2_k 
\left(\frac{a'}{a}h'+ h\frac{a^{\prime\prime}}{a}
-{h\over 4}\left(\frac{a'}{a}\right)^{2}+{h^{\prime
\prime}\over 2} - {k\over ah}\right)\nonumber\\
G_{55}&=&\frac{3}{4}\left(\left(\frac{a'}{a}\right)^{2}+\frac{h'}{h}
\frac{a'}{a} - 4{k\over ah}\right)
\eeqa

Next we will rewrite 
\beq
   a(r) = a_0 e^{-A(r)},
\eeq
where $a_0$ is an
arbitrary constant with dimensions of (length)$^2$,
and consider the following
linear combination of the Einstein tensor components: $2G_{00}/h +
2G_{11}/ (a \Sigma^2_k)$.  Using the Einstein equations,
$G_{mn}=\k^2\left(T_{mn}^{B}+T_{\mu \nu}^{b}\sqrt{h}\delta^{\mu}_{m}
\delta^{\nu}_{n}\delta(r-r_{0})\right)$, where $\delta^{\mu}_{r}\equiv 0$,
we obtain
\beqa
\label{int1.1}
&&(h'+hA')'-\frac{3}{2}A'(h'+hA') + 4{k\over a_0}e^A\nonumber\\
&&=2\k^{2}\left(\frac{T_{00}^{B}}{h}+\frac{T_{11}^{B}}{a\Sigma^2_k}\right)
+2\k^2\left(\frac{T_{00}^{b}}{h}+\frac{T_{11}^{b}}{a\Sigma^2_k}\right)
\sqrt{h}\delta(r-r_{0})\nonumber\\
\eeqa
This can be integrated once since $(h'+hA')'-\frac{3}{2}A'(h'+hA')$
is proportional to
$(e^{-3A/2}(h'+hA'))'$.  Using $Z_2$ symmetric boundary conditions at
the brane to interpret the contribution of the delta function, we have
\beqa
\label{int1.2}
\left(h'+hA')\right|_r&=&-2\k^{2}e^{\frac{3}{2}A}\int_{r}^{r_{0}}
\left(\frac{T_{00}^{B}}{h}+
\frac{T_{11}^{B}}{a\Sigma^2_k}\right)e^{-\frac{3}{2}A}dr\nonumber\\
&-&\k^{2}e^{\frac{3}{2}(A(r)-A(r_{0}))}
\left.\left(\frac{T_{00}^{b}}{h}+\frac{T_{11}^{b}}{a\Sigma^2_k}\right)
\sqrt{h}\right|_{r_{0}}\nonumber\\
&+& 4{k\over a_0}e^{\frac{3}{2}A}\int_r^{r_0} e^{-\frac{1}{2}A} dr.
\eeqa
Let us for the moment consider the cases of vanishing or negative spatial
curvature, $k=0$ or $-1$.  Then at the horizon $r=r_{H}$, $h=0$ and since
the right hand side of eq.\ (\ref{int1.2}) is not positive---assuming that
$-T_0^0 + T_1^1 \geq 0$, in accordance with positivity of the stress
energy tensor---we conclude that $h'\leq 0$ at the horizon. But the brane
is located at a value $r_0>r_h$ beyond the horizon, by construction, and
$h(r_{0})$ must be positive for $t$ to be a timelike coordinate.  This
implies that $h'(r_h)>0$ at the horizon, as shown in figure 1, and in
contradiction to eq.\ (\ref{int1.2}).

From the argument above, we can discern two ways of evading the no-go
result.  (1) Violate positivity of the stress-energy tensor in the bulk
or on the brane, which would change the sign of either of the first two
integrals in eq.\ (\ref{int1.2}).  (2) Let the 3-D curvature be positive,
making the third integral in (\ref{int1.2}) positive.  In section IV we
will consider the second possibility.  For the moment, let us assume that
neither (1) nor (2) is fulfilled, and moreover that the 3-D curvature is
zero, which is the most favorable case for horizon formation, apart from
the positive curvature case.

\bigskip
\centerline{\epsfxsize=3.5in\epsfbox{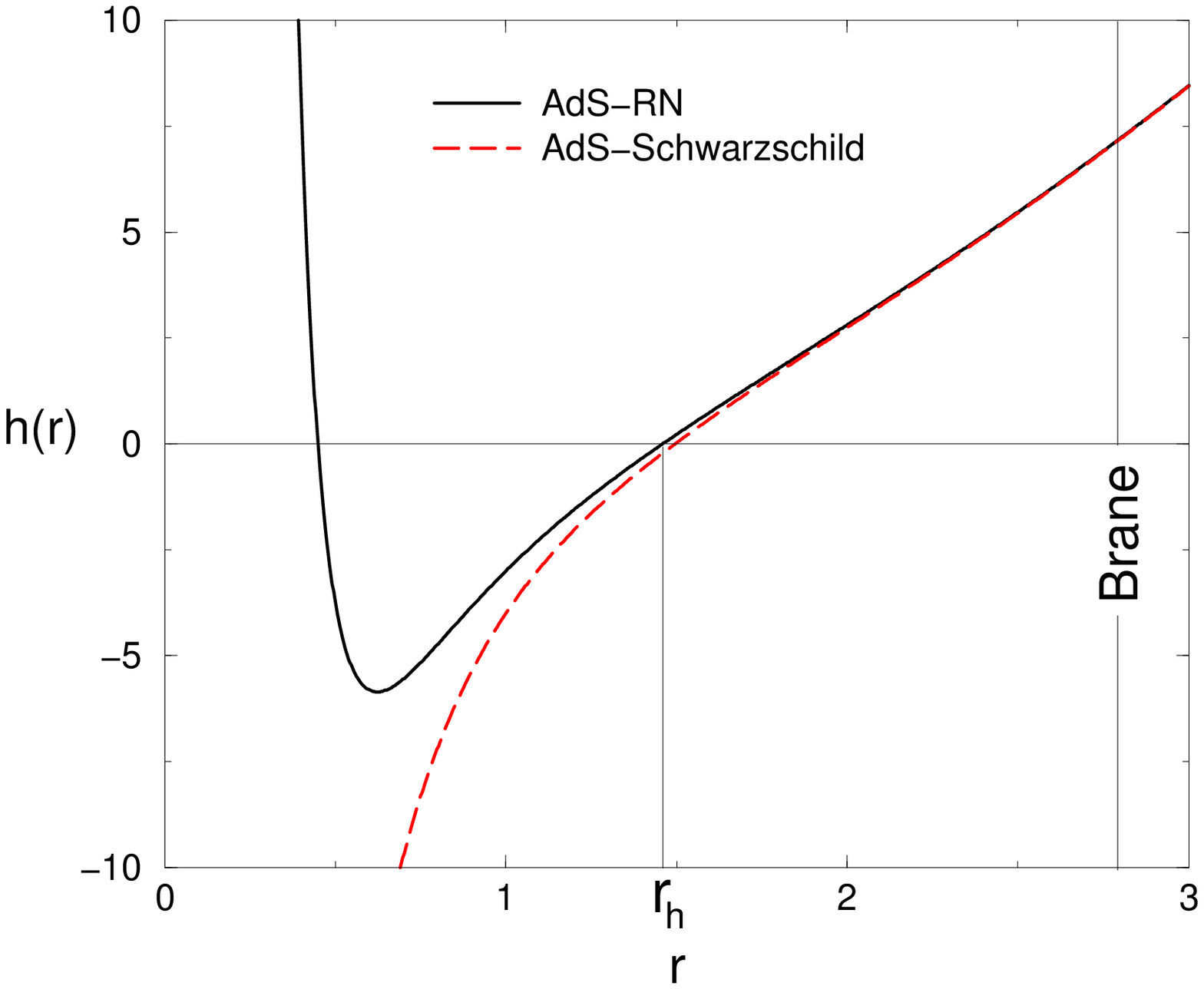}}
\vspace{-0.05in}
\noindent {\small
Figure 1: Required behavior for $h(r)$ near the horizon, $r=r_h$, for
charged and for neutral black holes.  Since the
brane is at values of $r$ greater than $r_h$ and $h$ is supposed to be
positive in this region, $h'(r_h)$ must be positive.}
\vspace{+0.1in} 

To demonstrate the usefulness of our no-go theorem, we will now study
two examples.  Inclusion of a scalar field in the bulk along with gravity
is the simplest one. In this case the stress-energy tensor contributions
are 
\beqa
T_{mn}^{B} &=& -g_{mn}V+\partial_{m}\Phi\partial_{n}\Phi
-\frac{1}{2}g_{mn}\partial^{l}\Phi\partial_{l}\Phi\nonumber\\
(T^{b})^{m} _{n} &=& {\rm diag}(-V_{0},-V_{0},-V_{0},-V_{0},0)
\eeqa
where $V$ and $V_0$ are the potentials in the bulk and on the brane,
respectively.  Static solutions which respect Lorentz symmetry on the
brane will be of the form $\Phi = \Phi(r)$, {\it i.e.,} dependent on the
bulk coordinate only.  Then from the above expressions we see that
$\frac{T_{00}}{h}+\frac{T_{11}}{a}=0$ both in the bulk and on the brane
and hence $(h'+hA')=0$ everywhere in the bulk, as a consequence of eq.\
(\ref{int1.2}).  In this case $h$ must be of the form $e^{-A}$, just like
the metric component $a$.  Thus the metric is explicitly Lorentz
invariant, regardless of the choice of bulk or brane potentials, and there
can be no horizon, with the possible exception of the usual AdS horizon at
$r=\infty$, where $h$ and $a$ vanish together.  In this case the situation
is the same as was investigated in ref.\ \cite{CEGH}, which found that
self-tuning solutions where gravity is localized have a naked singularity.

As a second example, we consider the AdS-RN solution which ref.\
\cite{CEG} investigated in depth.  Here one introduces a U(1) gauge field
in addition to the negative vacuum energy in the bulk, giving the
stress-energy tensor
\beq
T_{mn}^{B} =
-g_{mn}\Lambda-\frac{1}{4}g_{mn}F_{ab}F^{ab}+F_{mc}F_{\ n}^{c}
\eeq
The solution to the equations of motion is
\beqa
h(r)&=&\frac{r^{2}}{l^{2}}-\frac{\mu}{r^{2}}+\frac{Q^{2}}{r^{4}}
\nonumber \\ 
a(r)&=& r^{2} \nonumber\\
F_{tr}&=&\frac{\sqrt{6}Q}{\k r^{3}}
\eeqa
where $\mu$ and $Q$ are proportional to the black hole mass and charge,
respectively, and $l^{-2}=-\frac{1}{6}\k^{2}\Lambda$. Subtituting
this solution into the stress energy tensor we find
\beq
\frac{T_{00}^{B}}{h}+\frac{T_{11}^{B}}{a}=\frac{6Q^{2}}{\k^{2}r^{6}}>0
\eeq
From eq.\ (\ref{int1.2}) it is then clear that there is no possibility of
having a horizon unless the positivity of the stress energy tensor on the
brane is violated, {\it i.e.,} $\rho+p<0$, where $\rho= -(T^{b})^{\
0}_{0}$ and $p=(T^{b})^{\ 1}_{1}$.  We note in passing that the jump
conditions at the brane are
\beq
\label{jump}
	{[a']\over a} = -\frac23\kappa^2\rho\sqrt{h^{-1}};\qquad
	{[h']\over h} = \frac23\kappa^2(2\rho+3p)\sqrt{h^{-1}}
\eeq
where $[\ ']$ denotes the discontinuity in the derivative across the
brane.  Since $h'>0$ at the brane, its discontinuity assuming $Z_2$
symmetry is negative.  Hence the brane tension must be
positive; nevertheless $\rho+p$ is negative and the parameter $\omega
\equiv p/\rho$ must be less than $-1$.

It was recently proposed that adding a dilatonic coupling to the gauge
field will improve the situation in such a way that the horizon could be
outside of the brane \cite{GQTZ}, and the interior region containing the
singularity is cut away when the $Z_2$ symmetry around the brane is
imposed.  In this situation, to satisfy the jump conditions (\ref{jump}),
$\rho$ must be negative and $\omega\equiv p/\rho$ must be positive. This
is in contrast to the AdS-RN situation where $\omega < -1$ was required,
in contradiction to the positive energy condition.  The authors of ref.
\cite{GQTZ} find that $\omega>0$ in their new solution, which may at first
look like an improvement. However, when $\rho<0$, positivity of the
stress-energy tensor actually requires that $\omega \le -1$ (so that
$\rho+p>0$), so we see that the problem still persists in their 
solution.\footnote{
Ref.\ \cite{GQTZ} does however remark upon the possibility of overcoming
this problem when the curvature is $k=1$.}

In fact we can easily extend our no-go theorem to the case where the
brane is placed between the singularity and the horizon to show that no
improvement is provided by this variation.  Let us suppose there exists
a bulk solution which is qualitatively like one of those shown in figure
2; these are the negatives of the normal AdS-Schwarzschild or AdS-RN
solutions.  In this case the brane should not be placed at $r>r_h$ because
in this region $r$ is the timelike coordinate, and such a brane
would not be static, as we would like for self-tuning, but instead would
represent a time-dependent solution. Repeating the steps that led 
to eq.\ (\ref{int1.2}), we obtain
\beqa 
\label{int1.3}
\left.(h'+hA')\right|_r&=&2\k^{2}e^{\frac{3}{2}A}
\int_{r_{0}}^{r}\left(\frac{T_{00}^{B}}{h}+
\frac{T_{11}^{B}}{a\Sigma^2_k}\right)e^{-\frac{3}{2}A} dr\nonumber\\
&+&\k^{2}e^{\frac{3}{2}(A(r)-A(r_{0}))}
\left.\left(\frac{T_{00}^{b}}{h}+\frac{T_{11}^{b}}{a\Sigma^2_k}\right)
\sqrt{h}\right|_{r_{0}}
\nonumber\\ 
&-& 4{k\over a_0}e^{\frac{3}{2}A}\int_{r_0}^{r} e^{-\frac{1}{2}A} dr.
\eeqa

The new condition (\ref{int1.3}) is identical to the old one
(\ref{int1.2}) as far as the bulk contributions are concerned, but the
sign of the brane contribution is changed because of the fact that the
space is being cut away for $r<r_0$ rather than for $r>r_0$.  Now when we
apply (\ref{int1.3}) at the horizon, $r_h$, we find that $h'(r_h)$ gets
only positive contributions unless the 3-D curvature is $k=1$, or
positivity of $T_{\mu\nu}$ is violated.  But figure 2 makes clear that
$h'(r_h)$ should be negative in this case, thus giving a contradition.

\bigskip

\centerline{\epsfxsize=3.5in\epsfbox{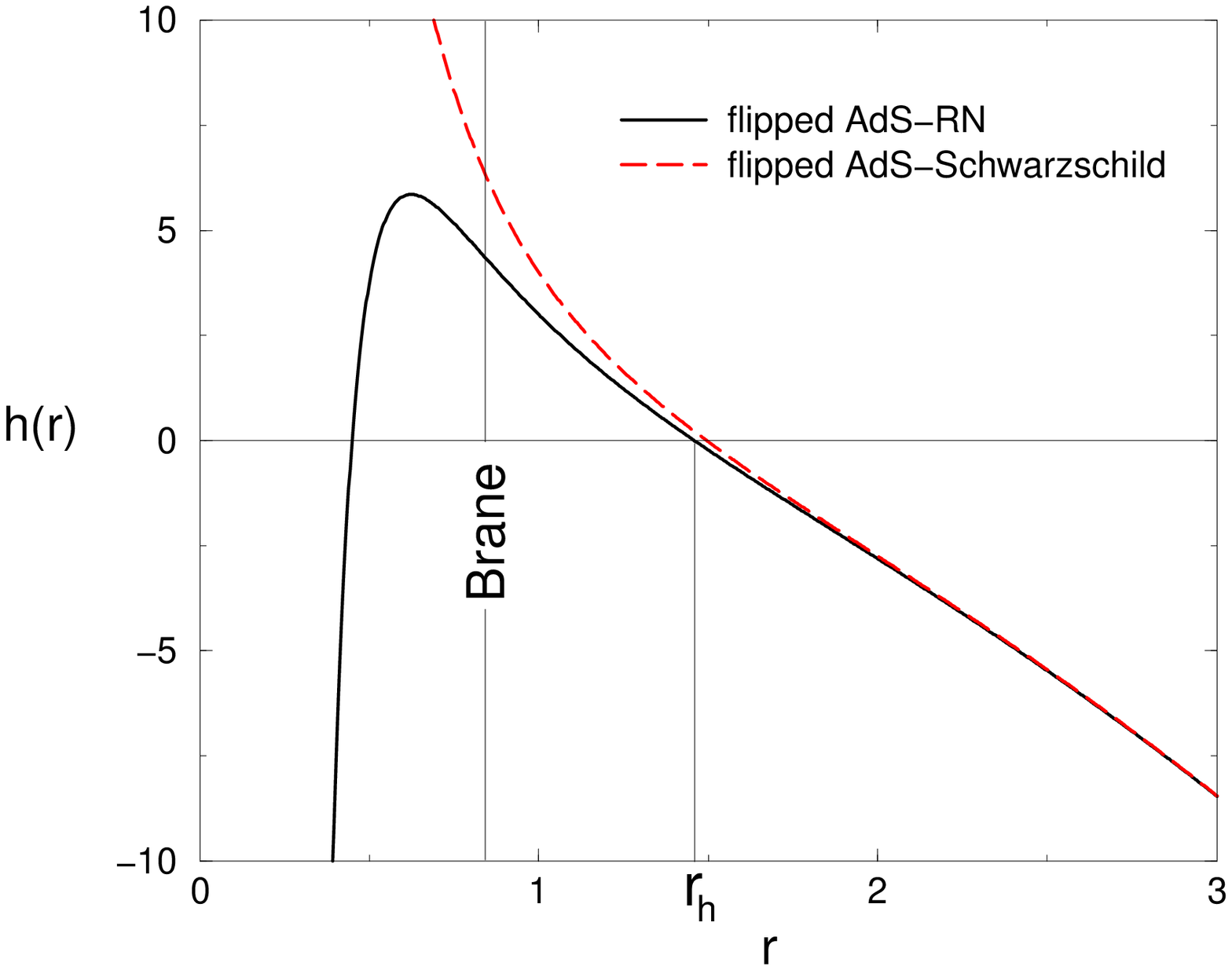}}
\vspace{-0.05in}
\noindent {\small
Figure 2.  Qualitative behavior of hypothetical solutions which would
require the brane to be placed between the horizon and the singularity.}
\vspace{+0.1in} 

\section{Relaxation of $Z_2$ Symmetry, and Higher Derivative Corrections}

We have seen that nonvanishing 3-D curvature $k$ can provide a way out of
our no-go result, but one may wonder whether there are other loopholes.  
In this section we continue to leave $k=0$ and explore two
possibilities which, as it
turns out, do not provide any additional loophole.  The first is to relax
the $Z_{2}$ symmetry imposed at the brane, at $r=r_0$.  We consider
possible solutions in which
\beq
	h(r) = \left\{\begin{array}{ll} h_1(r), & r< r_0\\
				    h_2(r), & r > r_0\end{array} \right. 
\eeq
and similarly for $a_i(r) = a_0 e^{-A_i(r)}$.  The solution would have
singularities on both sides of the brane, at positions $r=0$ and
$r=r_s>r_0$, say.  The hope would be to obtain horizons on both sides
before the singularities are reached, at $r=r_{h1},r_{h2}$, as illustrated
in figure 3.

\vspace{+0.1in} 
 Integrating eq.\ (\ref{int1.1}) in the bulk we get
\beqa
\label{int1.4}
\left.(h_{i}'+h_{i}A'_{i})\right|_r
&=&2\k^{2}e^{\frac{3}{2}A_{i}}\int_{r_{0}}^{r}
\left(\frac{T_{00}^{B}}{h_{i}}+
\frac{T_{11}^{B}}{a_{i}}\right)e^{-\frac{3}{2}A_{i}}dr\nonumber\\
&+&c_{i}e^{\frac{3}{2}A_{i}}
\eeqa 
where $c_{i}$ are the constants of integration determined by the jump
conditions (\ref{jump}).
Integrating eq.\ (\ref{int1.1}) across the brane gives
\beq
(c_{2}-c_{1})e^{\frac{3}{2}A(r_{0})}=\k^2\left.\left(\frac{T_{00}^{b}}{h}
+\frac{T_{11}^{b}}{a}\right)\sqrt{h}\right|_{r_{0}}
\eeq
 which implies $(c_{2}-c_{1})>0$.  But at the first horizon, $r_{h_1}$,
we need $h' >0$ and eq.\ (\ref{int1.4}) thus requires $c_{1}>0$. Similarly
getting $h' <0$ at $r_{h_2}$ requires $c_{2}<0$. These two conditions 
give rise to the contradictory relation $(c_{2}-c_{1})<0$; hence nothing
is gained by relaxing $Z_2$ symmetry.
 
\medskip
\centerline{\epsfxsize=3.5in\epsfbox{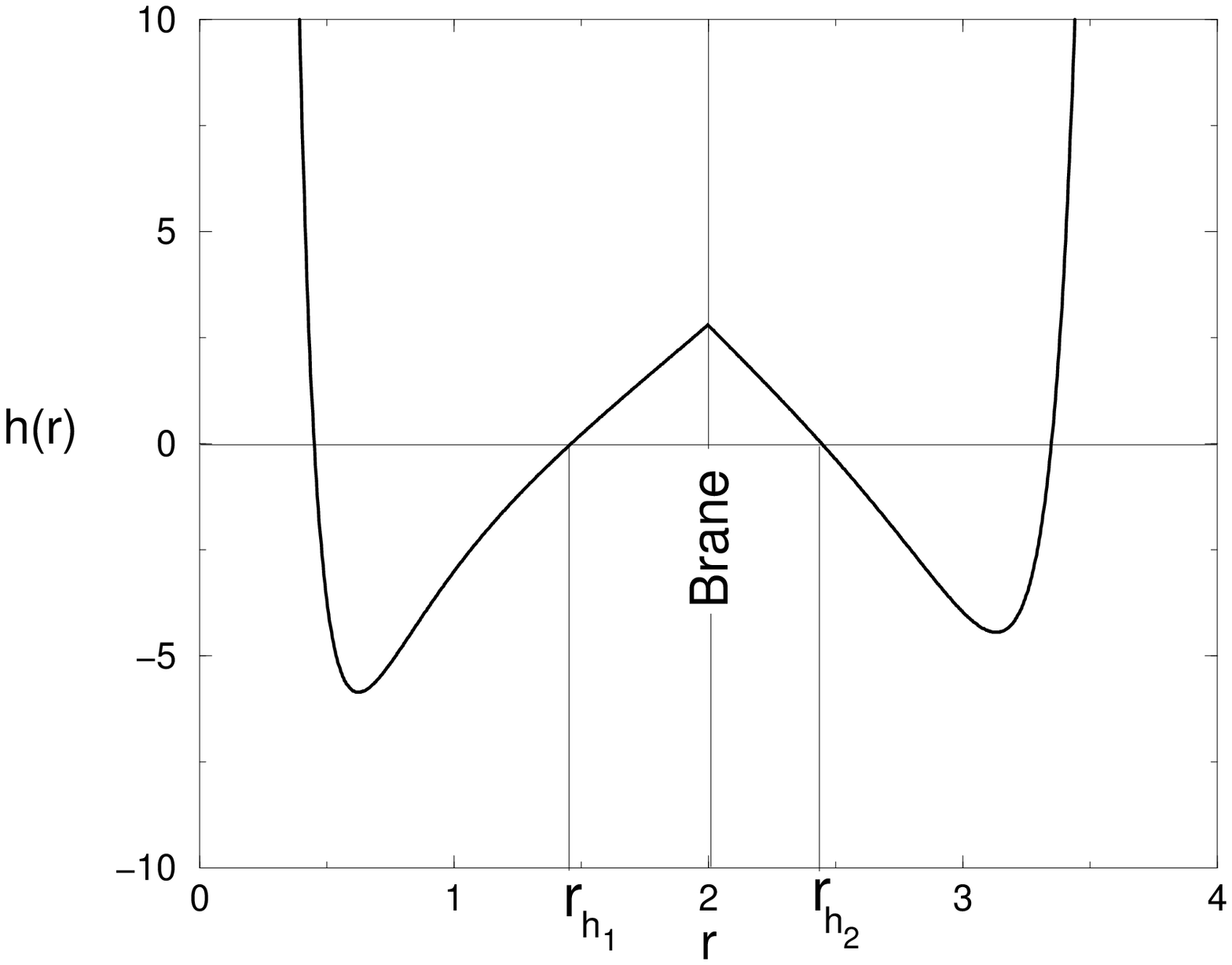}}
\vspace{-0.05in}
\noindent {\small
Figure 3.  Qualitative behavior of possible solutions without $Z_2$
symmetry across the brane.}
\medskip

One might alternatively hope that adding higher derivative corrections 
to the action might circumvent the no-go theorem. 
The simplest such correction to the Einstein-Hilbert action
is a Gauss-Bonnet term \cite{CB,CK}, since this introduces extra powers of
derivatives in the equations of motion without increasing the order of the
equations. The action is
\beq
S= \frac{1}{2\k^{2}}\int d^{\,5}x\sqrt{-g} \left(R+\lambda(R^{2}-4R_{ab}R^{ab}
+R_{abcd}R^{abcd}) \right)
\eeq
where $\lambda$ is the coefficient of the new Gauss-Bonnet term. The
modified Einstein equation is  ${\overline G_{mn}}=\k^{2}T_{mn}$  where 
${\overline G_{mn}}=G_{mn}+G^{(\lambda)}_{mn} $ and
\beqa
G^{(\lambda)}_{mn}&=&-\frac{\lambda}{2}g_{mn}(R^{2}-4R_{ab}R^{ab}
+R_{abcd}R^{abcd})\nonumber\\
  &+&2\lambda (RR_{mn}-2R_{mc}R^{c}_{n}+R_{mcde}R_{n}^{cde}+2R^{cd}R_{mcdn}) 
\nonumber\\
\eeqa
The (00) and (ii) components of the new contribution to Einstein's tensor are
\beqa
G^{(\lambda)}_{00}&=& +\frac{3}{4}h\lambda
\left(2\left(\frac{a'}{a}\right)^{2}
\frac{a^{\prime \prime}}{a}h^{2}
-\left(\frac{a'}{a}\right)^{4}h^{2}
+\left(\frac{a'}{a}\right)^{3}h'h\right)\nonumber
\\
G^{(\lambda)}_{ii}&=&-\frac{\lambda}{2}\left(2\frac{a^{\prime\prime}}{a}
\frac{a'}{a}-\left(\frac{a'}{a}\right)^{3}
+\left(\frac{a'}{a}\right)^{2}h'^{2}+
\left(\frac{a'}{a}\right)^{2}hh^{\prime\prime}\right)
\nonumber\\
\eeqa
As before, adding the (00) and (55) components of Einstein's equations
gives a differential equation similar to eq.\ (\ref{int1.1}), but
$(h'+hA')$ must be replaced with
$(h'+hA')(1-\lambda hA'^{2})$. Integrating this gives
\beqa
&&\left.(h'+hA')(1-\lambda hA'^{2})\right|_r = \nonumber\\
&&\qquad\qquad -2\k^{2} e^{\frac{3}{2}A}
\int_r^{r_0}
\left(\frac{T_{00}^{B}}{h}
+\frac{T_{11}^{B}}{a}\right)e^{-\frac{3}{2}A}dr' \nonumber\\
&&\qquad\qquad
\left.-\k^{2}e^{\frac{3}{2}(A(r)-A(r_{0}))}\left(\frac{T_{00}^{b}}{h}+
\frac{T_{11}^{b}}{a}\right)\sqrt{h}\right|_{r_0}.
\eeqa
The new factor does not change anything with respect to achieving a
horizon since $h$ vanishes there.  Just as before, the theorem shows that
at the putative horizon $h'<0$, in contradiction to the required behavior.

\section{Solutions with 3-D curvature}

In contrast to the negative results described above, it is possible to
have both self-tuning and a horizon, with no violations of stress-energy
positivity, when the spatial curvature is nonvanishing and positive.  In
fact, such a solution has already been obtained in ref.\ \cite{GJS} in the
case of the chargeless black hole with $k=1$.  The authors of \cite{GJS}
did not identify their solution as being self-tuning, but if one regards
the position of the brane along with the black hole mass as the properties
of the solution which adjust to compensate for the brane tension, then it
should indeed be considered as self-tuning. It is straightforward to apply
the jump conditions to show that a static solution with
\beq
	h(r) = k + {r^2\over l^2} - {\mu\over r^2};\qquad a(r)=r^2,
\eeq
exists if $k=1$, if the brane is placed at the position
satisfying
\beqa
	\kappa^4\rho^2 &=& {18\over r_0^2} + {36\over l^2}\nonumber\\
	\rho &>& 0;\qquad p = -\rho
\eeqa
(again using $l^{-2}\equiv -\frac16 \kappa^2\Lambda$) and if the black
hole mass parameter is
\beq
	\mu = \frac12 r_0^2.
\eeq
Thus self-tuning works for the range of brane tensions $\rho >
6\kappa^{-2} l^{-1}$.  The horizon is located at 
\beq
	r_h^2 = \frac12\left(-l^2 + \sqrt{l^4 + 2r_0^2l^2}\right)
\eeq
which can be shown to be always between the singularity and the brane.

We can easily generalize the above solution to the case of a charged black
hole, where
\beq
        h(r) = k + {r^2\over l^2} - {\mu\over r^2} + {Q^2\over r^4}
\eeq
The jump conditions determine the mass and charge of the black hole to 
be\footnote{We follow ref.\ \cite{CEG} in adopting the following
$Z_2$ parity assignments for the gauge field: $A_r\to + A_r$,
$A_t\to-A_t$, so that no new jump condition arises for it.}
\beqa
	\mu &=& r_0^4\left({3\over l^2} + {2\over r_0^2} -
	{\kappa^4\rho^2\over 12}\right)\nonumber\\
\label{muQ2}
	Q^2 &=& r_0^6\left({2\over l^2} + {1\over r_0^2} - 
        {\kappa^4\rho^2\over 18}\right)
\eeqa
The additional constant of integration, $Q$, introduces some freedom in
the position of the brane, which now can have a range of values for a
given brane tension.  The condition for existence of the horizon becomes
complicated because $h(r_h) = 0$ is a cubic equation.  It has real roots
(hence a horizon) only if the following inequality coming from the
discriminant of the cubic equation is satisfied:
\beq
\left({2\over 27}l^4 + \frac13\mu l^2 + Q^2\right)^2 < 
	{4\over 27} l^2\left(\frac13 l^2 + \mu\right)^3
\eeq
If we define
\beq
\label{params}
	\epsilon\equiv l^2/r_0^2;\qquad \eta \equiv
\frac{1}{36}\kappa^4\rho^2 l^2
\eeq
and use the expressions (\ref{muQ2}), this can be rewritten as $ [
1-\eta+\epsilon(1-\frac12\eta) + \frac13 \epsilon^2
+\frac{1}{27}\epsilon^3]^2 < [1 - \eta + \frac23\epsilon +
\frac19\epsilon^2]^3$.  Although this is hard to solve analytically, the
results are shown numerically in figure 4.  The darkened region is where
horizons exist for positive values of $Q^2$. The region above the
wedge ($\eta > 1 + \epsilon/2$)  corresponds to $Q^2 < 0$.  The lower
region has $Q^2$ exceeding the critical value beyond which the horizons
are lost, resulting in a naked singularity. The boundary of this region can be
approximated by the line $\eta \cong 1 + \frac{12}{27}\epsilon$ (the
approximation becoming exact as $\epsilon\to\infty$). Thus the allowed
region for self-tuning with a horizon is given approximately by
\beq
	1+\frac{12}{27}\epsilon \lsim \eta \le 1 + \frac12 \epsilon
\eeq
These solutions correspond to a brane with positive tension since the
discontinuity of $h'$ at $r_0$ is negative.

\bigskip
\centerline{\epsfxsize=3.5in\epsfbox{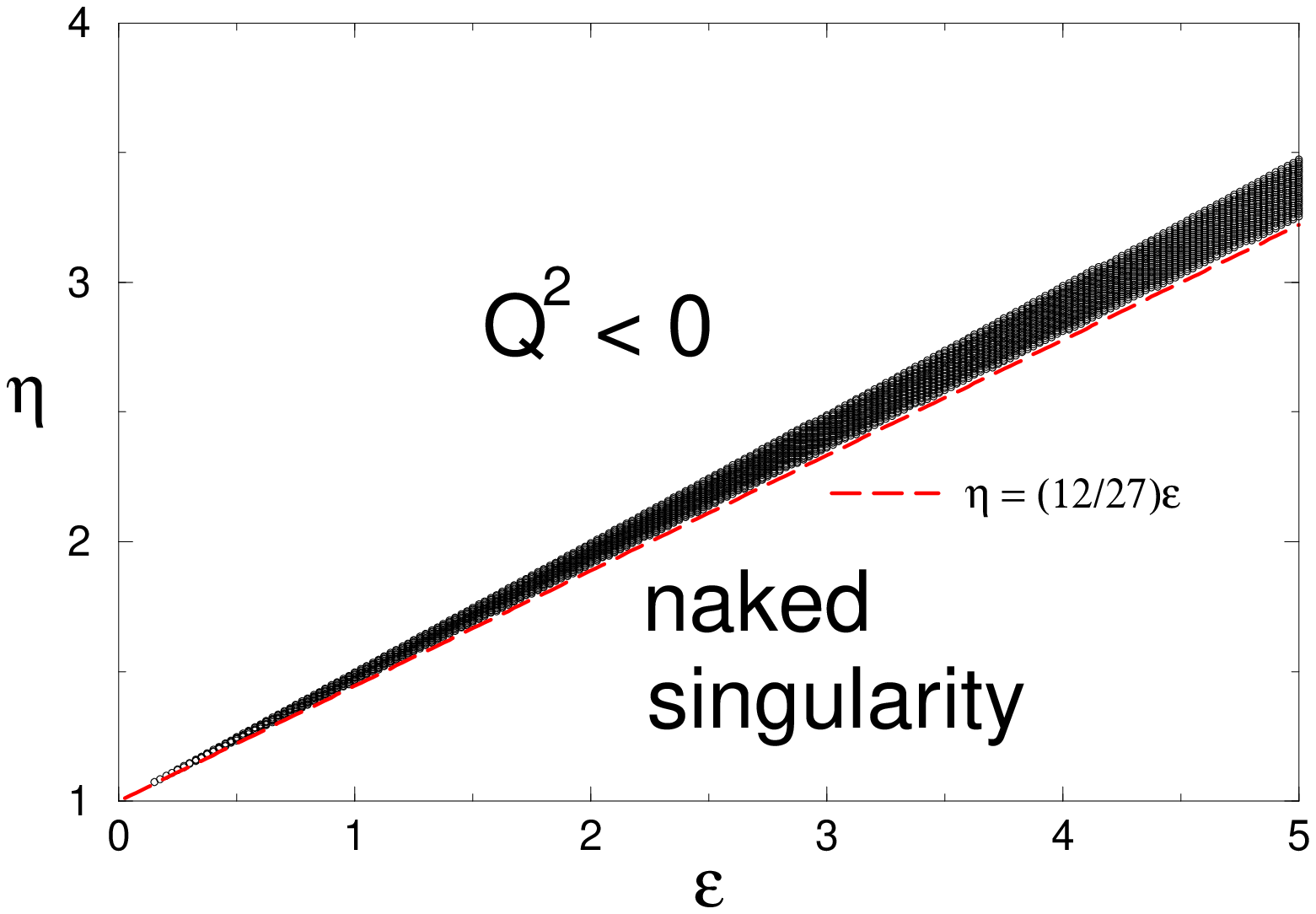}}
\noindent {\small
Figure 4: The dark wedge is the range of
parameters (see eq.\ (\ref{params})) for which self-tuning with a horizon
and with $Q^2 > 0$ occurs for the $k=+1$ AdS-RN solution.  The upper limit
corresponds to $Q=0$, and the lower one to the critical charge for which
the horizons disappear.}
\vspace{+0.1in} 

We thus see that generalizing to $Q^2 > 0$ does not significantly relax
the relation between the brane's tension and its position relative to the
original chargeless solution, unless $r_0\ll l$.  However the small $r_0$
regime is not very physical, because at distances much shorter than $l$,
the AdS curvature scale, one expects higher derivative corrections to the
gravitational action to alter the solution, so that one should not trust
it in detail for $r\ll l$.  Moreover, our universe would have to have very
large values of $r_0$ in order to be nearly spatially flat.

We have also searched for static solutions with negative tension branes
and $k=1$ in the AdS-RN case, where the brane is between the singularity
and the horizon.  These could in principle exist because of the inner
horizon and the positivity of $h(r)$ in this region.  However we do not
find any such solutions.  All those illustrated in figure 4 have positive
tension branes located outside of the horizons.  To arrive at this
conclusion, we numerically evaluated the positions of the horizons,
$r_{h_i}$, on a fine grid in the $\eta$-$\epsilon$ plane and checked
whether $r_0$ was less than or greater than these values.  The solutions
of $h(r_h)=0$ are given by
\beq
	{r^2_h\over l^2} = -\frac{1}{3} + 2\sqrt{A\over 3}\cos\theta
\eeq
where
\beqa
	A &=& \frac13 + {\mu \over l^2};\qquad B = 
	\frac{2}{27} + {\mu\over 3 l^2} + {Q^2\over l^4};\nonumber\\
	\theta &=& \frac13\left(n\pi + \tan^{-1}\sqrt{{4A^3\over 27B^2}-1} 
	\,\right),\quad n=0,1,\dots,5
\eeqa
(The three extraneous roots of the six given by this procedure were
identified by substituting back into the original equation.)  There are
always two roots with positive $r_h^2$ and one unphysical one with negative
$r_h^2$, except in the regions $\eta \ge 1 + \epsilon/2$ ($Q^2\le 0$)
(where there is only one horizon)  and $\eta \lsim 1+ (12/27)\epsilon$
(where $Q^2$ exceeds the critical value for having any horizon). 
The physical values of $r_h$ are always less than that of the
brane position, $r_0$.

\section{CURVED EXTRA DIMENSIONS}

Since we live in a universe that is nearly flat, it would be better if
the effect of a large brane tension could be counteracted by the curvature
of some small extra dimensions rather than that of the usual three.  To 
explore whether this is possible, we consider the following ansatz for the
metric, in which two extra dimensions with the geometry of a two-sphere
of radius $\sqrt{b(r)}$ are introduced:
\beqa
\label{metric2}
ds^2=&-&h(r)dt^{2}+a(r)d\Sigma_{0}^{2}+{h(r)}^{-1}dr^{2}\nonumber\\
&+&b(r)(d\theta^{2}
+\sin^{2}\theta d\phi^{2})
\eeqa
The (00) and (ii) components of the Einstein tensor are
\beqa
G_{00}&=&-\frac{3}{4}h\left(\frac{a'}{a}h'+2h\frac{a^{\prime
\prime}}{a}\right)\nonumber\\
&-&\frac{h}{4}\left(6\frac{a'}{a}\frac{b'}{b}h+2\frac{b'}{b}h'
-h\left(\frac{b'}{b}\right)^{2}+4\frac{b^{\prime\prime}}{b}h
-\frac{4}{b}\right)\nonumber\\
G_{ii}&=&a\left(\frac{a'}{a}h'+ h\frac{a^{\prime\prime}}{a}
-{h\over 4}\left(\frac{a'}{a}\right)^{2}+{h^{\prime
\prime}\over 2} \right)\nonumber\\
&+&\frac{a}{4}\left(4\frac{a'}{a}\frac{b'}{b}h+4\frac{b'}{b}h'
-h\left(\frac{b'}{b}\right)^{2}+4\frac{b^{\prime\prime}}{b}h
-\frac{4}{b}\right)\nonumber\\
\eeqa
Although the last terms in $G_{00}$ and $G_{ii}$ show the effect of the
positively curved extra dimensions, this effect cancels out of the
relevant linear combination $\frac{G_{00}}{h}+\frac{G_{ii}}{a}$.
Repeating the same steps that led to our previous no-go result gives
\beqa
&& \left(h'+hA')\right|_r=-2\frac{\k^{2}}{b}e^{\frac{3}{2}A}\int_{r}^{r_{0}}
b\left(\frac{T_{00}^{B}}{h}+
\frac{T_{11}^{B}}{a}\right)e^{-\frac{3}{2}A}dr\nonumber\\
&&\qquad-\k^{2}\frac{b(r_{0})}{b(r)}e^{\frac{3}{2}(A(r)-A(r_{0}))}
\left.\left(\frac{T_{00}^{b}}{h}+\frac{T_{11}^{b}}{a}\right)\sqrt{h}
\right|_{r_{0}}
\eeqa
Again we need $h'(r_{h})>0$, whereas the above expression shows that
$h'(r_{h})\leq 0 $, regardless of the curved extra dimensions.  The latter
thus do not provide any new loophole in our theorem.

\section{Discussion}

By integrating a certain linear combination of the (00) and (ii)
components of the Einstein equations, we have derived an enlightening
constraint on the $g_{00}$ component of the metric when there is a black
hole in a five-dimensional bulk, which is reminiscent of other consistency
conditions that have been deduced for brane/bulk solutions
\cite{sumrule}.  Our theorem explains why previous attempts to hide
self-tuning singularities behind a horizon have had to resort to a brane
equation of state which violates positivity of the stress-energy tensor.  
It shows that one could alternatively achieve the same effect by
violating positivity in the bulk rather than on the brane.  Indeed, in our
search for self-tuning solutions with a horizon and with $p=-\rho$ on the
brane, prior to deriving this theorem, we discovered that it is possible
if the black hole charge has unphysical values with $Q^2 < 0$.

In our quest for loopholes to this constraint, we found that augmenting
the gravitational action with higher powers of the curvature was
unsuccessful, as was relaxing the $Z_2$ orbifold symmetry that is often
assumed when cutting the bulk space off at the brane.  Including positive
spatial curvature for the 3-D hypersurfaces provided a more successful way
of evading the no-go theorem.  We noted that the previously discovered
solution of ref.\ \cite{GJS} was an example of self-tuning with a horizon,
and we generalized it by allowing the black hole to have a charge.

Unfortunately these positive curvature metrics do not provide a realistic
solution to the cosmological constant problem because the curvature is
related to the brane tension $\rho$ by $1/r_0 = C \sqrt{\rho}/M_p$, where
$C$ is a number which lies within a narrow range of values of order 1. In
our universe, which is nearly flat, the same relation exists, where $\rho$
is the critical energy density.  Therefore these ``self-tuning'' solutions
can describe our universe only if the brane tension is on the order of the
presently observed cosmological constant; this therefore constitutes
fine-tuning after all.  Precisely the same conclusion would hold if one
tried to counteract the effect of a positive cosmological constant with
positive curvature in a purely 4-D solution. This is nothing other than
Einstein's static solution, which is known to be unstable against
perturbations of the scale factor away from the special static value.  It
seems likely that the same problem will afflict the 5-D solutions as well. 
Moreover, we found that it was not possible to cancel the effect of the
brane tension by shifting the positive curvature into small extra dimensions. 
Despite these somewhat negative conclusions however, given the notorious
difficulty of the cosmological constant problem, the self-tuning approach
with a horizon nevertheless seems deserving of further exploration.

\bigskip
{\bf Note added}: the issue of shielding singularities in the bulk
by horizons, and the difficulties of so doing, have been previously
considered in ref.\ \cite{Gubser2}.  We thank S.\ Gubser for reminding
us about this work.

\bigskip

We thank Christophe Grojean, Csaba Csaki, and Joshua Erlich for very
helpful communications.  H.F.\ thanks Neil Constable for his kind
assistance with the Maple GRTensor package.  J.C. acknowledges the
generous hospitality of NORDITA during the completion of
this work.


\end{document}